\documentclass[12pt]{article}

\begin{document}

{\large The theory of GMR and TMR in segmented magnetic nanowires.}\\[5mm]

M. Ye. Zhuravlev$^{a,*)}$, H. O. Lutz$^a$, A. V. Vedyayev$^{b,**)}$\\[3mm]

$^a${\it Fakult\"{a}t f\"{u}r Physik, Universit\"{a}t Bielefeld,
33501 Bielefeld 1, Germany}\\
$^b${\it CEA/D\'{e}partement de Recherche Fondamentale sur la
Mati\'{e}re Condens\'{e}e, SP2M/NM, 38054 Grenoble, France}\\
$^{*)}${\it on leave from N. S. Kurnakov Institute of General and 
Inorganic Chemistry of
the RAS, 117907 Moscow, Leninskii prosp., 31, Russia}\\
$^{**)}${\it on leave from Department of Physics, Moscow State University,
Moscow, 119899, Russia}\\[5mm]


 We calculate the resistivity and Giant Magnetoresistance 
(GMR) of a segmented nanowire consisting of
two ferromagnetic segments separated by a thin paramagnetic spacer.
The quantization of the electron motion due to the small nanowire
cross-section is taken into account; $s-d$ electron scattering
gives rise to different mean free paths  
for spin-up and spin-down $s$-electrons.
The calculated resistivity and GMR oscillate  
as a function of nanowire cross-section due to the difference in
Fermi momenta of $d$-electrons with opposite spins. 
The GMR can reach values much higher than those which are obtained for
"wires" of infinite cross-section (i.e., a multilayer). 
Similarly we have calculated the Tunneling 
Magnetoresistance (TMR) by replacing the paramagnetic spacer  
with an insulator spacer.

\newpage

  Since its discovery, the Giant Magnetoresistance (GMR) effect in
  magnetic multilayers \cite{Baibich}
  has attracted considerable attention due to its fundamental
  interest as well as its application potential
(for a recent reference on the subject cf. \cite{Special}). In most 
studies so far the magnetic structures consist of a number of
thin layers of very large lateral extension; thus, quantum size
effects and the corresponding charge localisation occurs only in
the component perpendicular to the layers. It is to be expected,
however, that progress in miniaturisation of technical elements
will require the understanding of systems whose lateral extension is
limited, too. This is the topic of our letter. Specifically, we
investigate electron transport perpendicular to the layer
structure. In contrast to earlier theoretical studies [3-6] of
this CPP geometry ({\bf c}urrent {\bf p}erpendicular to {\bf
p}lanes of infinite lateral extension), we treat the electron
transport in a segmented nanowire: two ferromagnetic segments are
separated by a thin spacer, and charge localisation is thus
threedimensional. As a consequence, the GMR oscillates not only
as a function of layer thickness, but also as a function
of the wire cross section.

In the above-mentioned earlier studies of magnetic multilayers
[3-6] the free-electron model has been used taking into account
the exchange splitting of the d-bands, yielding different values
of the elastic mean free paths for up and down spin $s$-electrons.
More recently, the GMR was calculated employing realistic band
structures of the ferromagnetic and paramagnetic metallic layers
\cite{schep,butl,tsym,stil,math}. One of the main conclusion was
the observation that a correct description of the GMR can be 
achieved if $sp-d$ hybridization is taken into account.

In the present work we extend the quantum theory of the GMR
presented e.g. in ref.\cite{vedy4}; it is based on the theory of
spin-dependent scattering of conduction electrons as developed in
ref.\cite{vedy3}, assuming that $s$-electrons give the main
contribution to the current due to their low effective 
mass \cite{fert2} . Their mean free
path depends on the spin due to $s-d$ hybridization and the
different density of states (DOS) of the $d$-electrons at the
Fermi level. Comparable to the treatment of the layers as a random
binary alloy, we calculate the mean-free paths in the framework
of the coherent potential approximation(CPA), using the main
conclusion of \cite{brau}  in that the effective mean free paths
$l^{\uparrow, \downarrow}$ of the $s$-electrons are proportional
to the DOS $\rho_{d \uparrow, \downarrow}$ of the d-electrons (the
arrows indicate the spin direction). The conduction electrons are
treated as a free electron gas, and the details of the band
structure are neglected. We want to point out, however, that in
spite of the simplicity of this model we have taken $s-d$
hybridization into account.

The calculations of resistivities and magnetoresistance are
performed on the basis of the Kubo formalism in accordance with
the general scheme \cite{vedy4}. First we consider a sample with
rectangular cross section of size $a\times b$; the layer
thicknesses are $c_j$, with the layer index j=1 and 3 referring
to two outer ferromagnetic parts and j=2 to a paramagnetic 
spacer.

The one-electron Green function 
$G^{\sigma}(x,y,z,x',y',z')$ ($0<x,x'<a$, $0<y,y'<b$)
in coordinate-energy representation
is the solution of the following equation:
\begin{equation}
\left(\Delta+{\displaystyle
{2ME^{\sigma}_{j}\over\hbar^2}}\right)G^{\sigma}=
{\displaystyle{2Ma_0\over\hbar^2}}\delta(\vec{r}-\vec{r'}),
\label{puas}
\end{equation}
obeying the continuity of $G^{\sigma}$ and its derivatives at the
interfaces between the layers; in addition we impose the condition
$G^{\sigma}=0$ on the outer (lateral) interfaces. Index ${\sigma}$ 
denotes the spin projection of the electron.  
The complex value $E^{\sigma}_j$
depends on the layer $j$; in case of a paramagnetic spacer one
obtains
\begin{equation}
E^{\sigma}_j = {\displaystyle{2ME\over\hbar^2}} + 
(k^{\sigma}_{Fj})^2 +
i{2k^{\sigma}_{Fj}\over l^{\sigma}_{j}} \label{cpa}
\end{equation}
where $E$ is the energy relative to the 
Fermi energy, $l^{\sigma}_j$ is the
mean free path and $k^{\sigma}_{Fj}$ is the Fermi 
momentum of electrons with spin projection ${\sigma}$ in
the $j$-layer. The real part of the coherent potential is
included in the Fermi energy. Both the $d$- and the $s$- electron
Green functions obey equation (\ref{puas}) if the respective
 parameters are used. We suppose that the spin splitting of the
$s$-band is negligibly small. The solution of Eq. (\ref{puas}) can
be written in the form
\begin{equation}
G^{\sigma}(\vec{r},\vec{r'})=\sum\limits_{n,m}
{4\over ab}G^{\sigma}_{nm}(z,z')\sin{\pi nx\over a}\sin{\pi my\over b}
\sin{\pi nx'\over a}\sin{\pi my'\over b}
\end{equation}
where $G^{\sigma}_{nm}(z,z')$ obeys the equation
\begin{equation}
\left({{\partial}^2\over \partial z^2}+\epsilon_{nm}^{\sigma (j)}\right)
G^{\sigma}_{nm}(z,z')=
{\displaystyle{2Ma_0\over\hbar^2}}
\delta(z-z')
\label{z'z}
\end{equation}
with $$ \epsilon^{\sigma (j)}_{nm}= 
{\displaystyle{2ME^{\sigma}_j\over\hbar^2}}-
\left({\pi n\over a}\right)^2- \left({\pi m\over b}\right)^2 $$
Eq.(\ref{z'z}) coincides with the equation for the Green function
for infinite multilayers as obtained in \cite{vedy3}, were also
the solution of this equation has been presented. Note that the
values of the mean free path $l^{\sigma}_j$ and the 
Fermi momentum $k^{\sigma}_{Fj}$
depend on the size of the wire cross section; they can be
calculated in the following way:

To renormalize the Fermi energy we equate the total ($s$ and $d$)
electron concentration $n = 2n_s+n_{d\uparrow}+n_{d\downarrow}$
for an infinite volume to the concentration of the electrons in a
finite-size sample. Here $n_s$ is the one-half concentration of
$s$-electrons and $n_{d\sigma}$ is the concentration of
$d$-electrons with spin $\sigma$. The electron concentrations are
given by
\begin{equation}
n_{s,d\sigma} = -{\mbox Im}{1\over\pi}{1\over abc}
 \int\limits_0^{E_F}  { \int G^{\sigma}(\vec{r}, \vec{r}, E)d^3r}dE
\end{equation} .

The resulting $k^{\sigma}_{Fj}$ values are then used to calculate the mean
free path of the $s$-electrons: $$ {\displaystyle{l^{\sigma}(a,
b, c)\over l^{\sigma}(a\to\infty, b\to\infty, c\to\infty)}}=
{\displaystyle{\rho_{d\sigma}(a, b, c)\over
\rho_{d\sigma}(a\to\infty, b\to\infty, c\to\infty)}} . $$ Note
that both the density of states and the mean free paths of
s-electrons oscillate with different periods as a function of
nanowire cross section due to the different Fermi momenta of
d-electrons with opposite spins.

The current perpendicular to the  multilayer plane is given in the
framework of the Kubo formalism as
\begin{equation}
\begin{array}{l}
J(z)={1\over\pi}{e^2\over\hbar^2}
\left({\hbar^2\over2M}\right)^2
{\int}
\sum\limits_{n,m=1}^{\infty}
\Biggl\{\Bigl[{\partial G^{\sigma}_{nm}(z,z')\over \partial z}-
{\partial G^{\sigma *}_{nm}(z,z')\over \partial z}\Bigr]
\Bigl[{\partial G^{\sigma}_{nm}(z,z')\over \partial z'}-
{\partial G^{\sigma *}_{nm}(z,z')\over \partial z'}\Bigr]-\\
\phantom{123}
\Bigl[{\partial^2 G^{\sigma}_{nm}(z,z')\over\partial z \partial z'}-
{\partial^2 G^{\sigma *}_{nm}(z,z')\over\partial z' \partial z}\Bigr]
[G^{\sigma}_{nm}(z,z') - G^{\sigma *}_{nm}(z,z')]\Biggr\}E(z')dz'
\end{array}
\end{equation}

As in the case of laterally infinite multilayers \cite{vedy4} we
assume constant effective fields inside each layer to achieve
nondivergence conditions for the current $\partial J(z)/\partial
z=0$; the total voltage across the length of the nanowire
$U=\sum^3_{j=1}\mathcal{E}_j c_j$, with $\mathcal{E}_j$ the
effective field in the jth segment directed along the z-axis.

Next, the GMR can be calculated from the resistivities for
parallel ($R(\uparrow\uparrow)$) and antiparallel
($R(\uparrow\downarrow)$) magnitizations
 of the ferromagnetic layers:

\begin{equation}
{\displaystyle {\Delta R\over R}}=
{\displaystyle {R(\uparrow\downarrow) - R(\uparrow\uparrow)
\over {\mbox min}\{R(\uparrow\uparrow), R(\uparrow\downarrow)\}}}
\end{equation}

In addition, following the model described in ref.~\cite{lacroix},
we also calculated the tunneling magnetoresistance (TMR) of the
nanowire if the paramagnetic spacer is replaced by an insulating
spacer. In this case the spin-dependent resistance is due to the
$d$-electron behavior since the exchange splitting of the
$d$-band causes spin-dependent potential steps between the
segments. Green functions obey the same {\it Eqs.} (\ref{puas},2) with 
$k_{Fj}^{\sigma}$ replaced by $-V_0$, where $V_0$ 
is height of the barrier. Denoting the real and imaginary parts 
of the electron momentum as $k^{\sigma}_j$ and $d^{\sigma}_j$: 
$$
k^{\sigma}_j+id^{\sigma}_j=\sqrt{\epsilon^{\sigma (j)}_{nm}}
$$
one obtains for $d_2c_2 \gg 1$ an 
expression for the conductivity of electrons with 
spin projection $\sigma$ similar to $Eq.$(3) of
ref.~\cite{bratkovsky} but with discrete momenta:
\begin{equation}
\Sigma^{\sigma}= \sum\limits_{n,m}{e^2\over\pi\hbar}
{\displaystyle{16k_1^{\sigma}k_3^{\sigma}d_2^2\exp(-2d_2c_2)\over
((k_1^{\sigma})^2+(d_2+d_1^{\sigma})^2)
((k_3^{\sigma})^2+(d_2+d_3^{\sigma})^2)}} 
,\label{tun}
\end{equation}
where the $k's$ and $d's$ depend on the indices $n, m$, 
and $d_2=\sqrt{V_0}$.
In the general case ($d_2c_2$ not $\gg$1), this expression is much more
complicated.

A few examples of our calculations are shown in Figs.~1-5 \cite{comm}. 
In Fig.1 the dependencies of the volume-averaged
$s$- and $d$- electron DOS on the wire cross section size ($a=b$) are
presented for Fermi momenta fixed at its value for 
$a=b\to\infty$ (Fig. 1a) as well as  
 renormalized according to the procedure described above
(Fig.1b). The curves show pronounced
oscillations with periods equal to $\pi/k_{Fj}^{\sigma}$. 
The required renormalization
of the Fermi energy leads to a noticeable change of the structure of
the curves. The thin curve with large amplitude of oscillations
describes the DOS of $s$-electrons in the paramagnetic spacer,
whereas the the heavy curve with small amplitude describes the
oscillations of the $d$-electron DOS in the ferromagnetic segments. The
large amplitude of the oscillations in the spacer compared with the 
amplitude of oscillations in the ferromagnetics is due to the small 
thickness ($c_2$) of the spacer. The dependence of $s$-electron  
conductivity on the size $a$ is defined by the oscillating behaviour
of the mean free paths and DOS of $s$-electrons. These dependencies 
have to be considered as a complicated superposition
of oscillating $s$ and spin up and spin down
$d$-electron's densities of states. They are shown in Fig. 2  
for ferromagnetic and antiferromagnetic configurations.
The parameters were chosen to reproduce
the maximum experimental GMR for a CPP geometry at 4.2 K
for $Co|Cu|Co$ multilayers \cite{pratt}.

Fig.3 shows the dependence of the GMR on the wire dimension $a$.
As in Fig. 2, this curve is the superposition of several
oscillating curves with different periods. In the maxima, the curve
reaches much higher values than in case of
$a = b{\rightarrow}{\infty}$, i.e., in a multilayer sample. To
interprete this enhancement we have to take into account that the
GMR is controlled by the spin polarization of the current which
(for negligibly small thickness of the paramagnetic spacer) is
proportional to the ratio 
$
\left\{(l^{\uparrow}-l^{\downarrow})/
(l^{\uparrow}+l^{\downarrow})\right\}^2 ;
$
since $l^{\uparrow}$ and
$l^{\downarrow}$ oscillate with different periods this ratio
oscillates as well and thus can produce high peaks, while for
$k_{F}^{\uparrow}{\approx}k_{F}^{\downarrow}$ the GMR will be almost
constant. As expected, the GMR oscillation becomes more
pronounced for decreasing nanowire dimensions. This is in
qualitative agreement with the experiment \cite{garcia} in which
GMR values between 40\% and 200\% were observed in few-atom sized
nanocontacts at 4K. In these experiments the contact sizes were
not controlled and vary statistically; therefore, the GMR values
were statistically dispersed as well. It was observed, however,
that the GMR is higher for samples with larger contact
resistance, e.g. smaller dimension. This observation confirms our
suggestion that the GMR can achieve high values in case of very
thin wires, possibly of the order of several 100 \% for dimensions in the
range of a few Angstroms. We also note that for some specific
Fermi momenta and mean free paths the calculated GMR changes sign
in some region of cross section size (Fig.4).

The calculated TMR is represented in Fig.5 for different
thicknesses of the insulating spacer. Firstly, we note that the
absolute value of the TMR as well as the form of the curves
strongly depend on the thickness of the spacer.
Secondly, the oscillation amplitude as a function of
  size $a$ is much smaller if compared with the GMR oscillations
 in a nanowire of the same size. The reason is that due to
the exponential factor (see Eq. (\ref{tun})) the conductivity is
mainly determined by the term with minimal $d_2 c_2$ ($n=m=1$).

Finally, we want to point out that the resistivities and the GMR
have been calculated for ideal outer interfaces and specular
reflection of the electrons. A roughness of the interfaces can
lead to a smoothing of the oscillations. This effect will be
addressed in a future publication.\\

 A. V. Vedyayev acknowledges the CENG DRMC SP2M for hospitality
and the Russian Foundation of Fundamental Research for financial
support.
 M. Ye. Zhuravlev is grateful to Bielefeld University
for hospitality. The work has been supported by the Deutsche
Forschungsgemeinschaft in the "Forschergruppe
Nanometer-Schichtsysteme".


\newpage
{\large Figure captions}\\

Fig.1a,b.\\
Normalized DOS  of $s$-electrons in the paramagnetic spacer (thin line) 
and $d$-electrons in the ferromagnetic segment (heavy line) 
with fixed bulk Fermi momentum (1a) and renormalized
Fermi momentum (1b). $k_{F}^{\uparrow}$=1.40 \AA$^{-1}$, 
$k_{F}^{\downarrow}$=0.40 \AA$^{-1}$, $k_{F}^s$=1.20 \AA$^{-1}$; 
$l^{\uparrow}$=15.0 \AA,
$l^{\downarrow}$=120.0 \AA, $l_2$=215.0 \AA; $c_1=c_3$=22.0 \AA, 
$c_2$=7.0 \AA. Here $\uparrow$ refers to electrons with spin parallel
to magnetization, $\downarrow$ refers to electrons with spin antiparallel to
magnetization  and $s$ refers to electrons in the paramagnetic
spacer. $c_j$ is the thickness of the corresponding layer $j$. \\

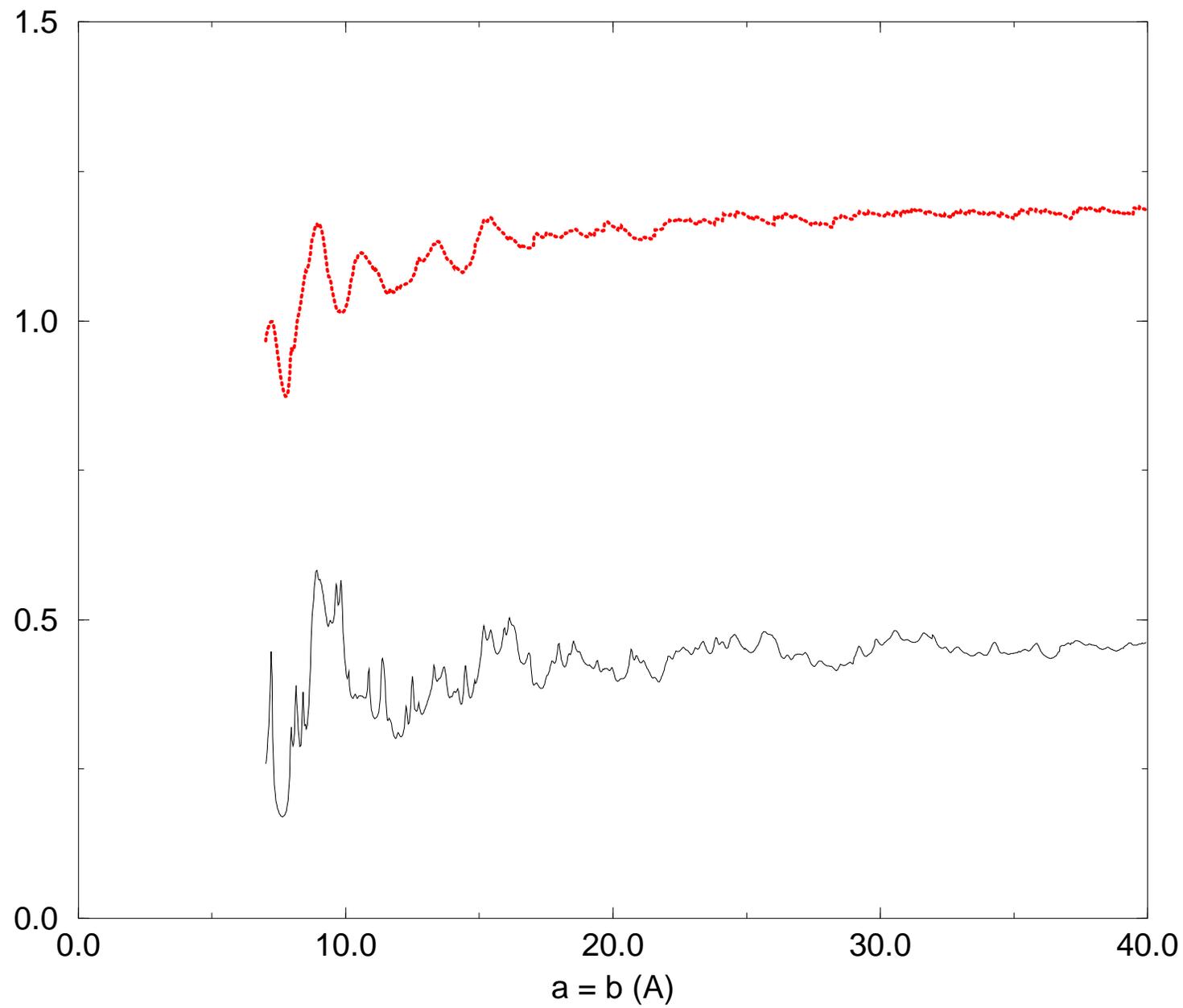
Fig.2. \\
The normalized resistivity for parallel magnetization,
$R(\uparrow\uparrow)$ (solid line)
and antiparallel magnetization, $R(\uparrow\downarrow)$ (dotted line) for
$k_{F}^{\uparrow}$=1.40 \AA$^{-1}$, $k_{F}^{\downarrow}$=0.40 \AA$^{-1}$, 
$k_{F}^s$=1.20 \AA$^{-1}$; $l^{\uparrow}$=15.0 \AA, 
$l^{\downarrow}$=120.0 \AA, $l_2$=215.0 \AA;
$c_1=c_3$=22.0 \AA, $c_2$=7.0 \AA.\\

Fig.3a,b. \\
GMR for the parameters used in Fig.1\\

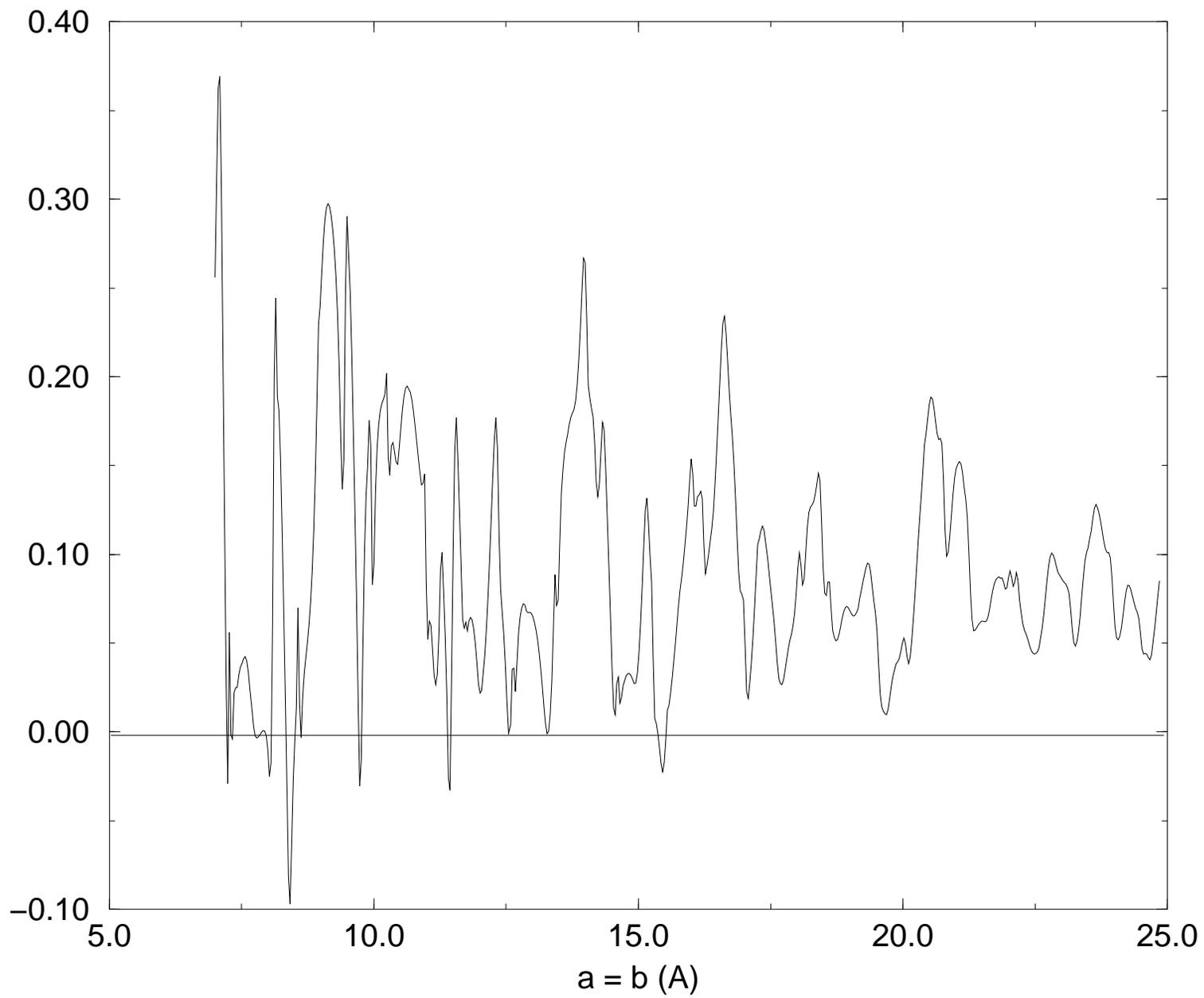
Fig.4.\\
GMR for $k_{F}^{\uparrow}$=0.87 \AA$^{-1}$, $k_{F}^{\downarrow}$=0.60
\AA$^{-1}$, 
$k_{F}^s$=1.19 \AA$^{-1}$; $l^{\uparrow}$=67.02 \AA, 
$l^{\downarrow}$=120.02 \AA,
$l_2$=215.02 \AA; $c_1$=22.8 \AA, $c_3$=23.2 \AA, $c_2$=7.3 \AA.\\

Fig.5a,b.\\
TMR for $k_{F}^{\uparrow}$=1.40 \AA$^{-1}$, 
$k_{F}^{\downarrow}$=0.40 \AA$^{-1}$, 
$V_0$=1.44 \AA$^{-2}$; $l^{\uparrow}$=21.0 \AA, $l^{\downarrow}$=100.0 \AA; 
the thickness of the ferromagnetic segments $c_1=c_3=$ 23.0 \AA, the thickness
of the insulator spacer $c_2=$ 4.0 \AA (a) and 20.0 \AA (b).


\begin{thebibliography}{99}

\bibitem{Baibich}
M.N. Baibich, Phys. Rev. Lett. {\bf 61} (1988) 2472; G. Binasch,
Phys. Rev. {\bf B39} (1989) 4828.

\bibitem{Special}
Special Issue: J. of Magentism and Magnetic Materials, Vol. 200,
No. 1-3 (1999).

\bibitem{fert1}
T. Valet and A. Fert, Phys. Rev. {\bf B 48} (1993) 7099;

\bibitem{levy2}
P. M. Levy, S. Zhang and A. Fert, Phys. Rev. Lett {\bf 65} (1990) 1643
H. E. Camblong, P. M. Levy and S. Zhang, Phys. Rev. {\bf B 51} (1995) 16052

\bibitem{vedy4}
A. Vedyayev, N. Ryzhanova, B. Dieny, P. Dauguet, P. Gandit ,
J. Chaussy, Phys. Rev. {\bf B 55} (1997) 3728

\bibitem{levy1} H. E. Camblong, S. Zhang and P. M. Levy,
Phys. Rev. {\bf B 47} (1993) 4735

\bibitem{schep} K. M. Schep, P. J. Kelly, and G. E. W.
Bauer, Phys. Rev. {\bf B 57}, (1998) 8907

\bibitem{butl}
W. H. Butler, X. G. Zhang, D. M. C. Nicholson, and J. M.
Maclaren, Phys. Rev. {\bf B 52} (1995) 13399

\bibitem{tsym} E. Yu. Tsymbal and D. G. Pettifor, Phys.
Rev. {\bf B 54} (1996) 15314

\bibitem{stil}
M. D. Stiles, J. Appl. Phys. {\bf 79} (1996) 5805

\bibitem{math}
J. Mathon, Phys. Rev. {\bf B 55} (1997) 960

\bibitem{vedy3}
A. Vedyayev, C. Cowache, N. Ryzhanova and B. Dieny,
J. Phys.: Condens. Matter {\bf 5} (1993) 8289

\bibitem{fert2}
A. Fert and I. A. Campbell, J. Phys. F: Met., {\bf 6} (1976) 849

\bibitem{brau}
F. Brouers, A. Vedyayev, and M. Giorgino, Phys. Rev. {\bf B 7}
(1973) 380

\bibitem{lacroix}
A. Vedyayev, N. Ryzhanova, C. Lacroix, L. Giacomoni and B. Dieny,
Europhys. Lett. {\bf 39} (1997) 219

\bibitem{bratkovsky}
A. M. Bratkovsky, Phys. Rev. {\bf B 56} (1997) 2344

\bibitem{comm}
{\small
In the calculations we supposed that the Fermi momenta in
both ferromagnetics depend only on the spin projection of the 
electron relative to the magnetization. Therefore, we use two 
parameters $k_{F}^{\uparrow,\downarrow}$ for the Fermi 
momenta of $d$-electrons with direction of spin 
parallel and antiparallel to the magnetization,  
and $k_{F}^s$ for the Fermi momenta of $s$-electrons in all
three segments. For simplicity we suppose that the Fermi momentum
of $d$-electrons in the paramagnetic segment equals to $k_{F}^s$ as well.
The same notations are held
for the mean free paths of $d$-electrons.
The bulk values of the corresponding Fermi momenta and 
the mean free paths are given in the
figure captions.
}


\bibitem{pratt}
W.P. Pratt, Jr., S.-F. Lee, J. M. Slaughter, R. Loloee,
P. A. Schroeder, and J. Bass, Phys. Rev. Lett,
{\bf 66} (1991) 3060;\\
P. Holody, W. C. Chiany, R. Loloee, J. Bass,
W.P. Pratt, Jr., and P. A. Schroeder,
Phys. Rev. {\bf B 58} (1998) 12230

\bibitem{garcia}
N. Garc\'{i}a, M. Mu\~{n}oz, Y.-W. Zhao, Phys. Rev. Lett.
{\bf 82} (1999) 2923


\end{thebibliography}
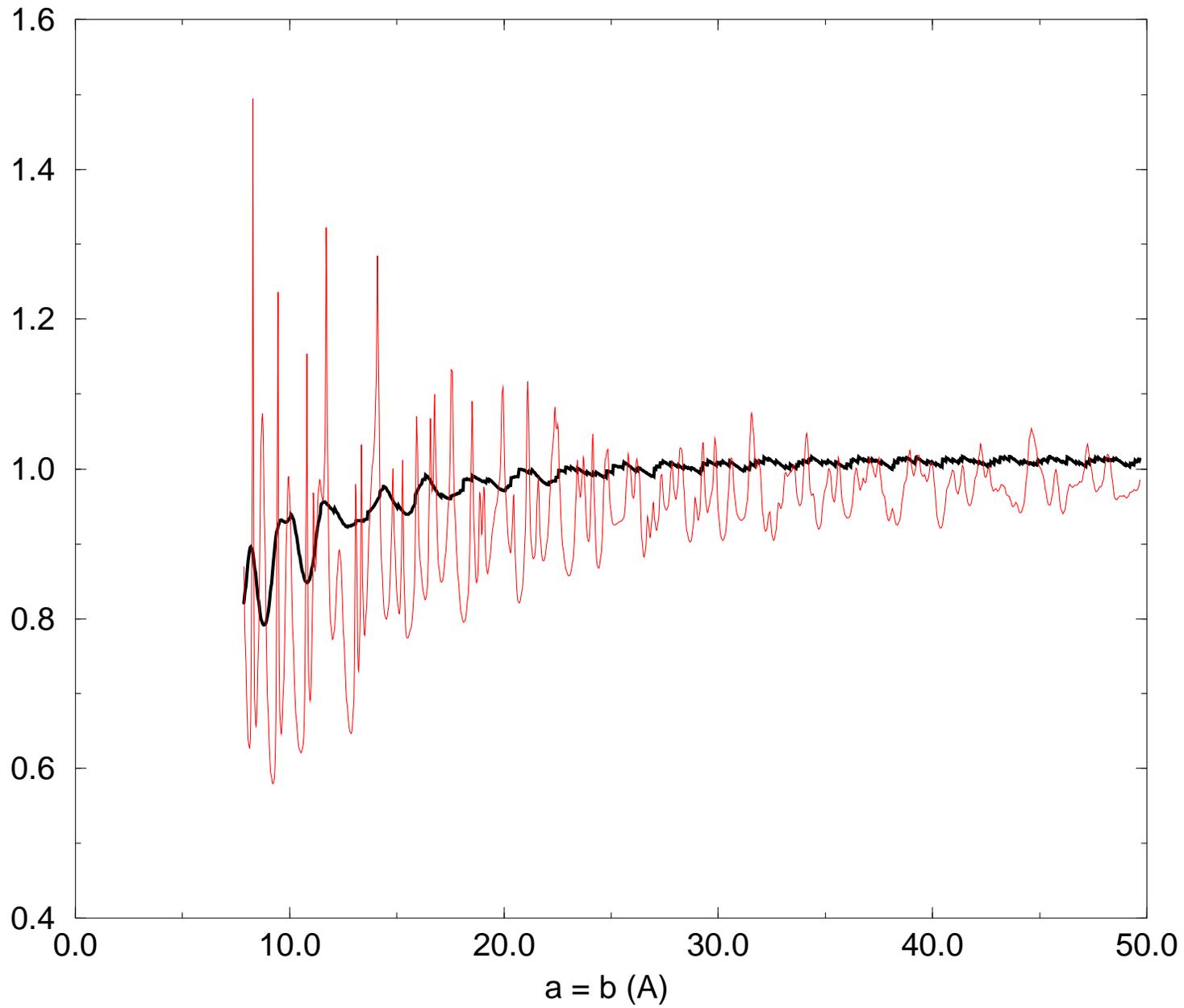
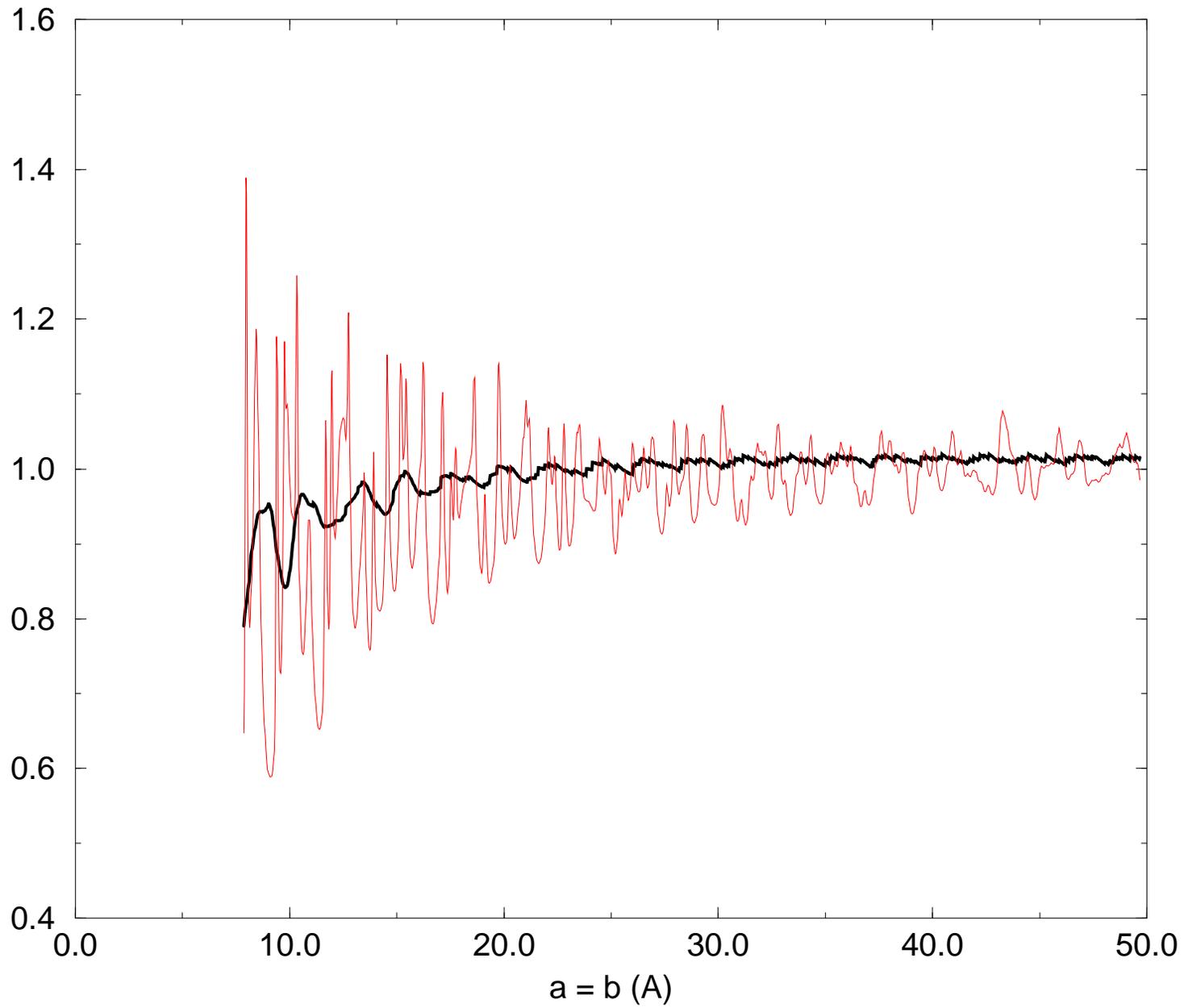
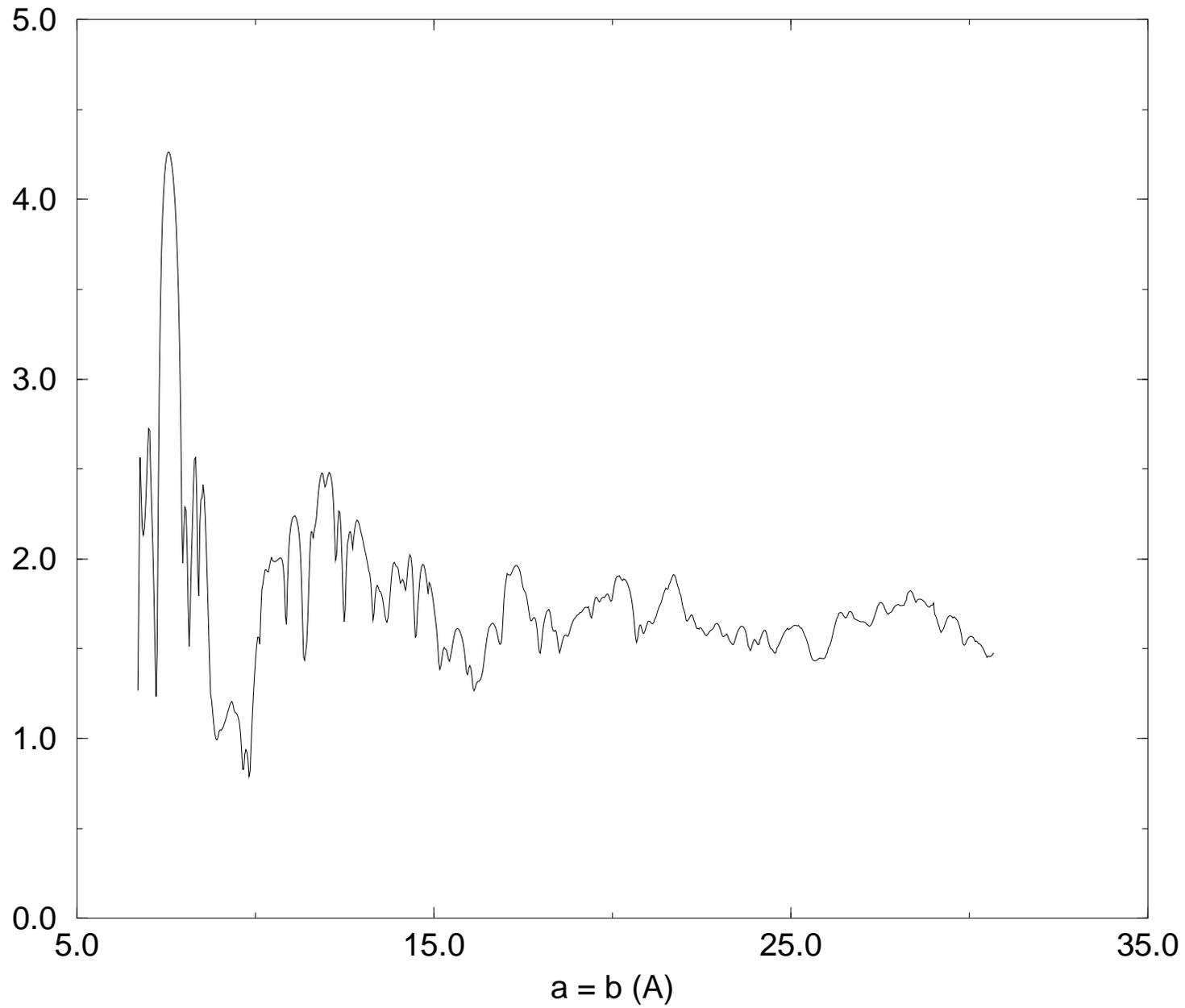
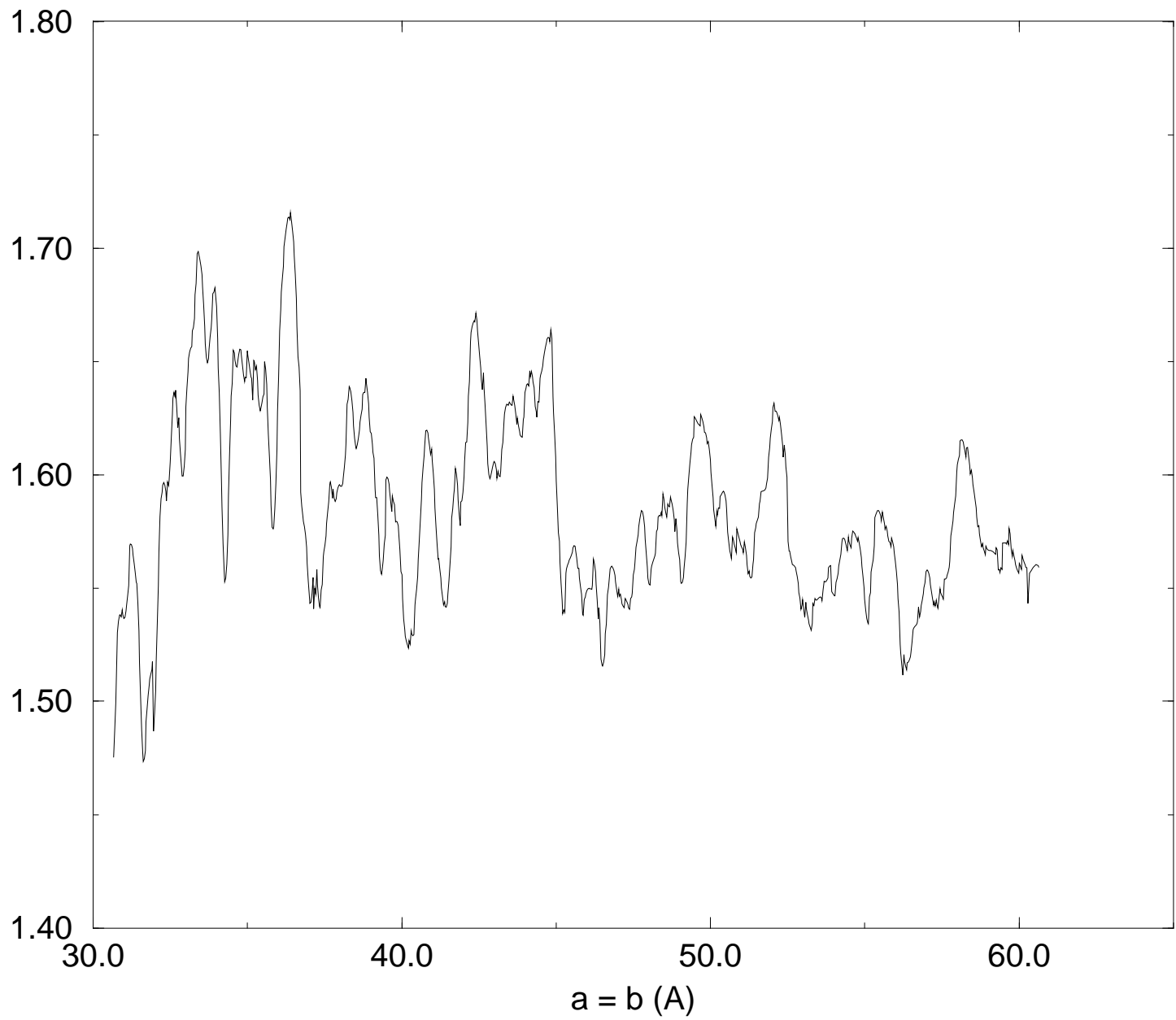
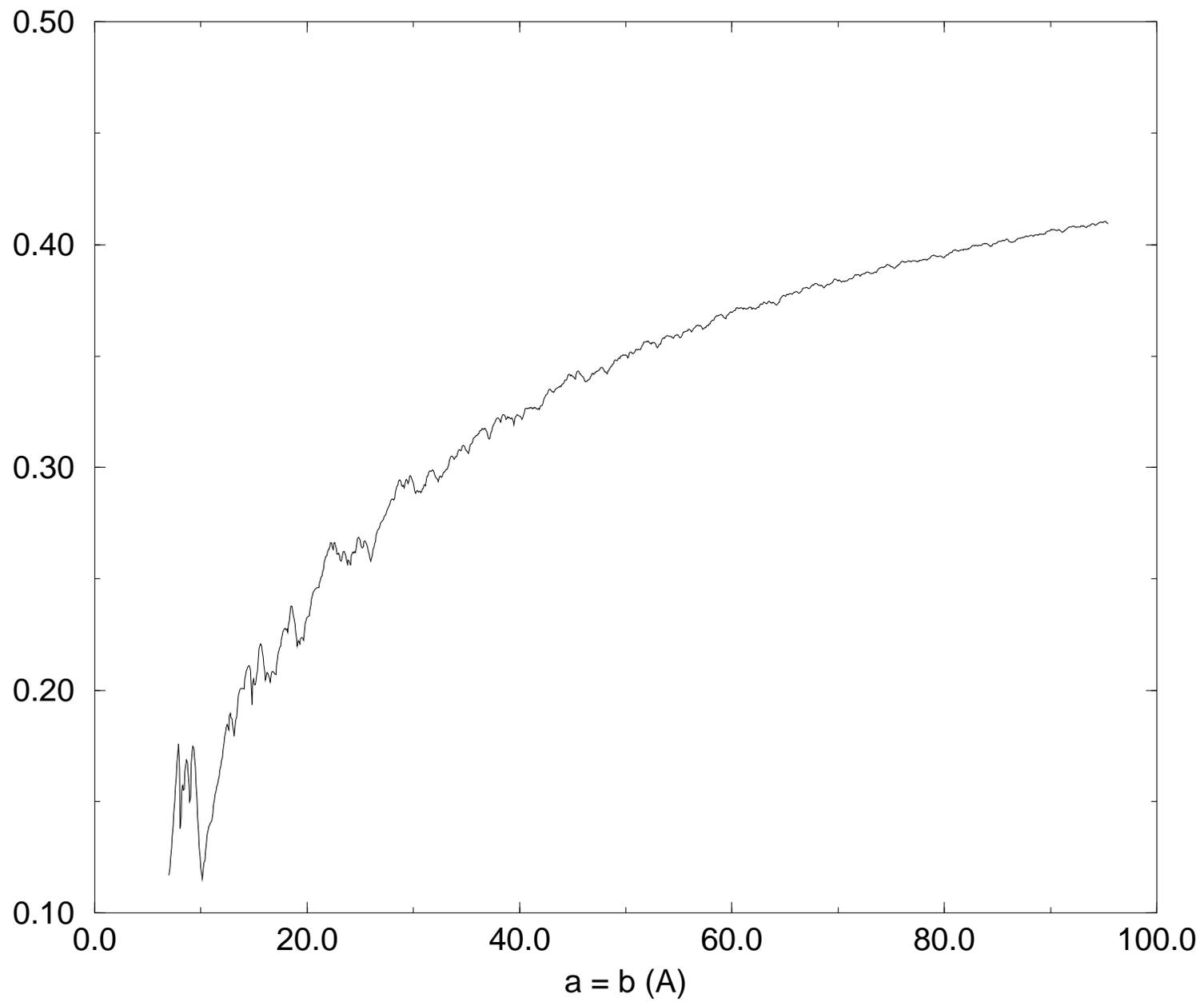
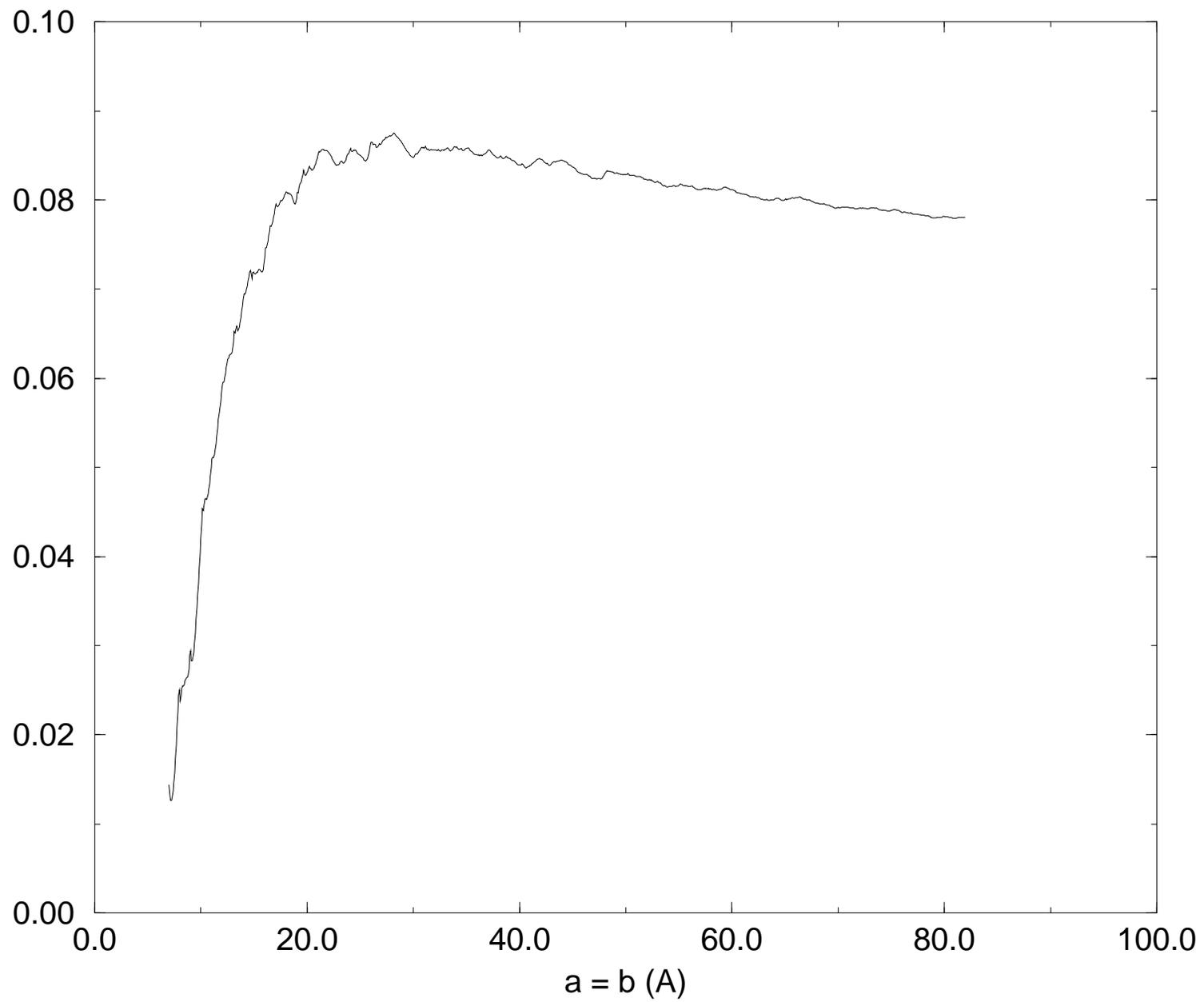
\end{document}